\documentclass[11pt]{article}

\usepackage[T1]{fontenc}
\usepackage{tcolorbox}

\usepackage{soul} 

\usepackage{garamond}  
\renewcommand{\rmdefault}{ugm}

\usepackage[bitstream-charter]{mathdesign}
\usepackage{amsmath}
\usepackage[scaled=0.92]{PTSans}

\usepackage[
  paper  = letterpaper,
  left   = 1.65in,
  right  = 1.65in,
  top    = 1.0in,
  bottom = 1.0in,
  ]{geometry}

\usepackage[final,expansion=alltext]{microtype}
\usepackage[english]{babel}
\usepackage[parfill]{parskip}
\usepackage{afterpage}
\usepackage{framed}

{\endMakeFramed}

\usepackage{lineno}

\usepackage{ragged2e}

\newcounter{parcount}

\usepackage{graphicx}
\usepackage[labelfont=bf]{caption}

\usepackage{booktabs}
\usepackage{multirow}

\usepackage[algoruled]{algorithm2e}
\usepackage{listings}
\usepackage{fancyvrb}
\fvset{fontsize=\normalsize}

\usepackage{natbib}

\usepackage[acronym,nowarn]{glossaries}
\makeglossaries

\definecolor{tangerine}{rgb}{0.95, 0.52, 0.0}
\definecolor{palebrown}{rgb}{0.6, 0.46, 0.33}
\definecolor{peru}{rgb}{0.8, 0.52, 0.25}
\usepackage[colorlinks,linktoc=all]{hyperref}
\usepackage[all]{hypcap}
\hypersetup{citecolor=peru}
\hypersetup{linkcolor=peru}
\hypersetup{urlcolor=peru}

\usepackage[nameinlink]{cleveref}
\creflabelformat{equation}{#1#2#3}
\crefname{equation}{eq.}{eqs.}  
\Crefname{equation}{Eq.}{Eqs.}

\lstdefinestyle{mystyle}{
    commentstyle=\color{OliveGreen},
    keywordstyle=\color{BurntOrange},
    numberstyle=\tiny\color{black!60},
    stringstyle=\color{MidnightBlue},
    basicstyle=\ttfamily,
    breakatwhitespace=false,
    breaklines=true,
    captionpos=b,
    keepspaces=true,
    numbers=left,
    numbersep=5pt,
    showspaces=false,
    showstringspaces=false,
    showtabs=false,
    tabsize=2
}
\usepackage[colorinlistoftodos,
           prependcaption,
           textsize=small,
           backgroundcolor=yellow,
           linecolor=lightgray,
           bordercolor=lightgray]{todonotes}

\usepackage{amsthm}  

\usepackage{bm}

\usepackage[colorinlistoftodos,
           prependcaption,
           textsize=small,
           backgroundcolor=yellow,
           linecolor=lightgray,
           bordercolor=lightgray]{todonotes}

\usepackage{graphicx}
\usepackage{caption}
\usepackage{subcaption}
\usepackage{wrapfig}

\usepackage{booktabs}
\usepackage{arydshln} 
\usepackage{multirow}
\usepackage{nicematrix}

\usepackage{listings}
\usepackage{fancyvrb}
\fvset{fontsize=\normalsize}

\usepackage[nameinlink]{cleveref}
\creflabelformat{equation}{#1#2#3}
\crefname{equation}{eq.}{eqs.}  
\Crefname{equation}{Eq.}{Eqs.}

\usepackage{listings}
\lstdefinestyle{alp_style}{
    commentstyle=\color{OliveGreen},
    numberstyle=\tiny\color{black!60},
    stringstyle=\color{BrickRed},
    basicstyle=\ttfamily\scriptsize,
    breakatwhitespace=false,
    breaklines=true,
    captionpos=b,
    keepspaces=true,
    numbers=none,
    numbersep=5pt,
    showspaces=false,
    showstringspaces=false,
    showtabs=false,
    tabsize=2
}

\usepackage{capt-of}

\usepackage{lipsum}

\theoremstyle{remark}
\newtheorem*{lemma*}{Lemma}

\usepackage{authblk}
\usepackage{enumitem} 
\usepackage{algorithmic}
\usepackage{tabularx}
\usepackage{booktabs}
\usepackage{pifont}
\usepackage[T1]{fontenc}
\usepackage{multirow}
\usepackage{threeparttable}

\title{\textbf{MuAPBEK: An Improved Analytical Kinetic Energy Density Functional for Quantum Chemistry}}

\author[1, 3]{Siwoo Lee}
\author[2, 3]{Adji Bousso Dieng}
\affil[1]{Department of Chemical \& Biological Engineering, Princeton University}
\affil[2]{Department of Computer Science, Princeton University}
\affil[3]{\href{https://vertaix.princeton.edu/}{Vertaix}}

\begin{document}
\maketitle

\begin{abstract}
\noindent Orbital-free density functional theory (OFDFT) offers a true realization of the Hohenberg-Kohn theorems, enabling full quantum-mechanical studies of electronic systems based solely on electron densities. However, OFDFT remains limited by the difficulty of formulating accurate kinetic-energy density functionals. In this paper, we substantially enhance the accuracy of OFDFT energies and densities by tuning, during density initialization, the parameter $\mu$ of the APBEK functional, which arises in the second-order gradient expansion of the kinetic energy for semiclassical neutral atoms. We augment this parameterized APBEK functional with two physically motivated, non-empirical corrections derived from Kato's cusp condition and the virial theorem. The resulting functional, which we call MuAPBEK, is benchmarked against Kohn–Sham density functional theory (KSDFT) on atoms, organic molecules from the QM9 dataset, and the anti-malarial drug artemisinin. MuAPBEK achieves much lower energy errors than standard APBEK and Thomas-Fermi-von-Weizs\"{a}cker functionals, even when the latter two are evaluated on converged KSDFT densities. Its mean absolute energy errors on atoms and molecules are $161$ and $122$ kcal/mol, respectively, indicating that MuAPBEK's errors do not scale with system size. MuAPBEK also yields accurate densities, with a mean integrated absolute density error of $1.8$ electrons for molecules. Importantly, one step of our density optimization scheme is at least ten times faster than a single KSDFT self-consistent field cycle and exhibits a lower-order computational time complexity of $\mathcal{O}(N^{1.96})$ with respect to system size, $N$. Our results indicate that highly-accurate OFDFT for large-scale quantum simulations beyond the practical limits of KSDFT is within reach.\\

\end{abstract}

Density functional theory (DFT), as formalized by the Hohenberg-Kohn theorems~\citep{hohenberg1964inhomogeneous}, establishes the existence of a universal functional, $E$, that maps an arbitrary electronic system's ground-state electron density, $\rho$, to its ground-state energy. This is expressed as 
\begin{equation}
\label{eqn: total_energy_functional}
    E[\rho] = \int V(\textbf{r}) \rho(\textbf{r}) d^3 \textbf{r} + T[\rho] + U[\rho]
\end{equation}
where $\int V(\textbf{r}) \rho(\textbf{r}) d^3 \textbf{r}$ is the exactly-solvable external potential energy, and $T[\rho]$ and $U[\rho]$ are universal functionals for the kinetic energy and electron-electron interaction energies, respectively. Thus, in principle, dramatically simplified studies of electronic systems is possible with $\sim$$\mathcal{O}(N)$ linear computational time complexity in system size, $N$, as the functionals solely depend on real-space density distributions. However, in practice, achieving chemical accuracy of $\sim$1 kcal/mol within this framework of orbital-free DFT (OFDFT) is a punishingly difficult task~\citep{witt2018orbital, ligneres2005introduction, wesolowski2013recent, xu2024recent}. This is primarily because an approximate $T[\rho]$ must be highly accurate, given that a system's kinetic energy makes up half of its total energy contributions. In the absence of an accurate kinetic energy functional, Kohn-Sham DFT (KSDFT) has become the dominant workhorse for quantum mechanical simulations of molecules and materials~\citep{kohn1965self, jones2015density}. KSDFT introduces fictitious Kohn-Sham orbitals to calculate kinetic energies and scales as $\mathcal{O}(N^3)$ due to orthonormalization of the orbitals and diagonalization of the Kohn-Sham Hamiltonian matrix. This becomes prohibitively expensive for systems larger than few hundreds of atoms even with supercomputer access. Clearly, the development of a highly-accurate kinetic energy functional would unleash accelerated explorations of the currently inaccessible vast chemical and material spaces \citep{ramakrishnan2017machine, von2020exploring, walsh2015quest}.

\begin{figure}[t]
\resizebox{\columnwidth}{!}{\includegraphics{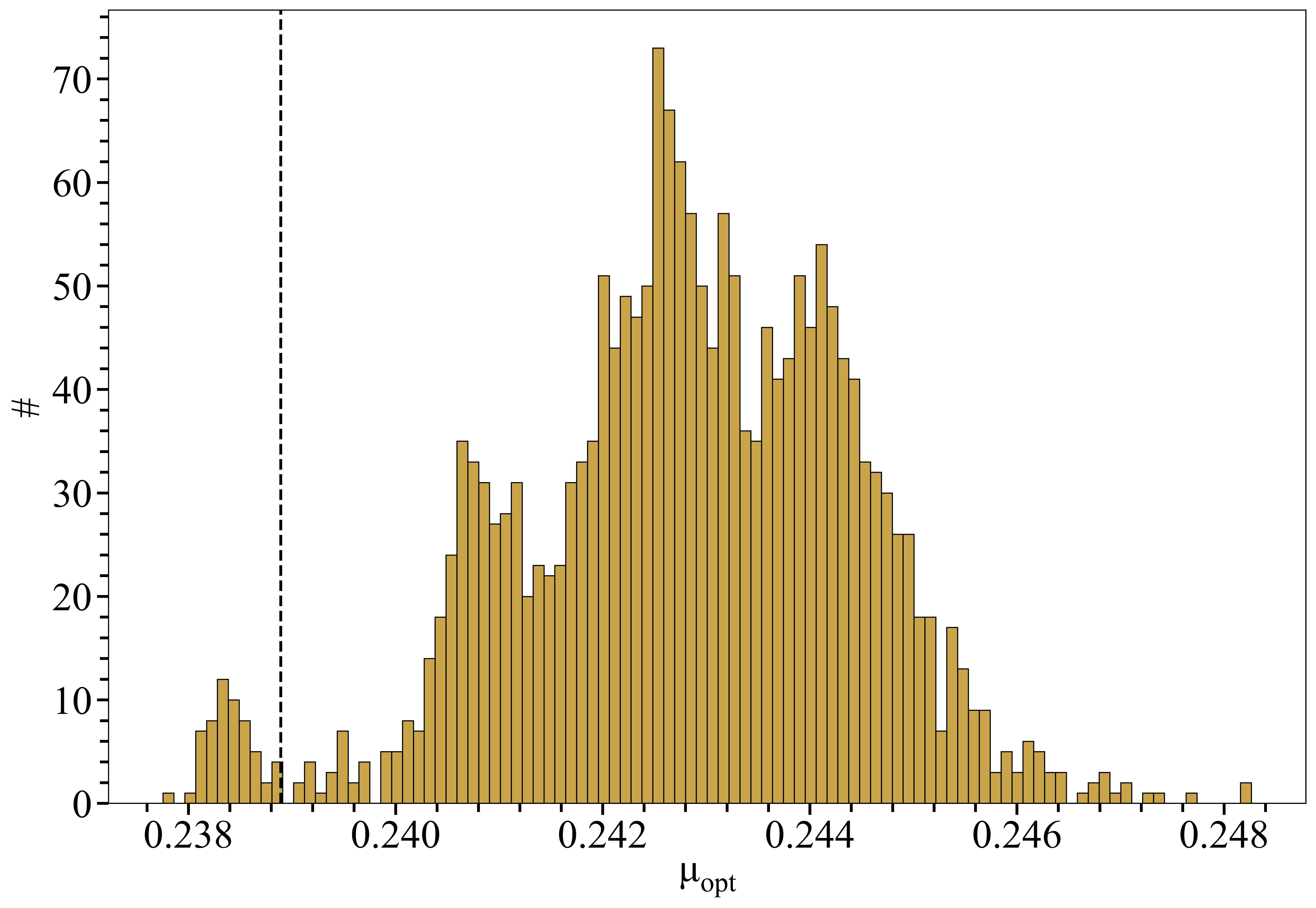}} 
\caption{\label{fig: Mu_opt_distribution}
Distribution of $\mu_{\textnormal{opt}}$ values for 2000 random QM9 test molecules computed with electron densities from converged PBE/def2-qzvp KSDFT calculations. 
The $\mu_{\textnormal{opt}}$ mean value of 0.24280 is 1.64\% larger than the default value of $\mu=0.23889$, indicated with the vertical dashed line. 
}
\end{figure} 

In this work, we consider the non-empirical generalized gradient approximation APBEK functional~\citep{constantin2011semiclassical} as a baseline for functional development. APBEK has shown encouraging results on a small molecular test set. It is derived from the asymptotic expansions of the kinetic energies of semiclassical neutral atoms and the conjointness conjecture~\citep{lee1991conjoint, tran2002link} which posits that the enhancement factor for a kinetic energy functional resembles that of an exchange-correlation functional. APBEK adopts the enhancement factor, $F$, from the Perdew-Burke-Ernzerhof (PBE) exchange functional~\citep{perdew1996generalized},
\begin{equation} 
\label{eqn: PBE_exchange_enhancement_factor}
    F(s) = 1 + \kappa - \dfrac{\kappa}{1 + \mu s^2 / \kappa}
\end{equation}
where $s = \lvert \nabla \rho(\textbf{r}) \rvert / \left( 2(3\pi^2)^{1/3} \rho(\textbf{r})^{4/3} \right)$ is the reduced density gradient. The parameters $\kappa \approx 0.80400$ and $\mu \approx 0.23889$ are respectively set to satisfy the Lieb-Oxford inequality~\citep{lieb1979lower, lieb1981improved} and the asymptotic second-order gradient expansions of semiclassical neutral atoms' kinetic energies.

We hypothesize that, while fixing $\kappa \approx 0.80400$ to satisfy the Lieb-Oxford inequality, $\mu$ can be made system-dependent to yield more accurate total energies. We first verify this by randomly selecting $2000$ molecules from the QM9 dataset~\citep{ramakrishnan2014quantum}, which contains organic drug-like molecules composed of up to $9$ heavy atoms. We compute their ground state energies and densities at the PBE/def2-qzvp~\citep{weigend2005balanced} level of restricted-spin KSDFT implemented in \texttt{PySCF}~\citep{sun2018pyscf}. Real-space grid point coordinates and weights are generated with Treutler-Ahlrichs radial grids~\citep{treutler1995efficient} and Becke partitioning~\citep{becke1988multicenter}. The optimal value of $\mu$, $\mu_{\textnormal{opt}}$, for each molecule at its ground state density is evaluated by minimizing the difference between the KSDFT and the APBEK total kinetic energy,
\begin{equation}
\label{eqn: mu_optimization}
    \mu_{\textnormal{opt}} = \arg \min_{\mu} \lvert T_{\textnormal{KSDFT}} - T_{\textnormal{APBEK}} \rvert
\end{equation}
where 
\begin{equation}
\label{eqn: APBEK_total_kinetic_energy}
    T_{\textnormal{APBEK}} = \int F \cdot \left( \dfrac{3}{10} \right) \left( \dfrac{6\pi^2}{2} \right)^{2/3} \rho(\textbf{r})^{5/3} d^3 \textbf{r} 
\end{equation} 
Performing this minimization yields the distribution of $\mu_{\textnormal{opt}}$ values in \autoref{fig: Mu_opt_distribution}, which has a mean value of $\mu_{\textnormal{opt}} \approx 0.24280$. 
Although this is a small increase in value of 1.64\% relative to the default one, it reproduces the exact energies and, therefore, highlights the importance of adapting $\mu$ to the system under study. Therefore, we identify two pressing goals that must be accomplished for APBEK to be used as or serve as a starting-point for an accurate standalone OFDFT functional: an accurate estimate of $\mu_{\textnormal{opt}}$, and the predictions of accurate densities and energies.

We tackle the first aim as follows. As with any KSDFT calculation, an initial guess of a system's electron density is made. We represent a spin-paired density as $\rho(\textbf{r}) = \sum_{i=1}^M \textbf{p}_{i} \omega_i(\textbf{r})$ where $\omega_i (\textbf{r})$ is the $i$-th basis function in an atomic density basis set with $M$ basis functions. To obtain the vector of density coefficients, $\textbf{p}$, the MINimal Atomic Orbitals (MINAO) initialization method~\citep{almlof1982principles, van2006starting} is used to form an initial guess of the density matrix as $\bm{\Gamma}_{\alpha \beta} = \sum_{i=1}^N \textbf{C}_{\alpha i} \textbf{C}_{\beta i}$, for the orbital coefficients $\textbf{C}_{\alpha i}$ where $\alpha, \beta$ index the atomic orbitals. 
The atomic orbital function values and their gradients, $\bm{\Phi}$, is evaluated on the real-space grid points generated as previously described. Density fitting~\citep{dunlap2000robust0, dunlap2000robust1} is performed with auxiliary basis sets fit to exchange and Coulomb potential terms~\citep{weigend2008hartree}. Then, the three-center two-electron repulsion integrals are calculated with the auxiliary basis functions as 
\begin{equation}
\label{eqn: three_center_AO_integrals}
    (\alpha \beta \vert L) = \int\int \phi_{\alpha}(\textbf{r}_1) \phi_{\beta}(\textbf{r}_2) \dfrac{1}{\lvert \textbf{r}_1 - \textbf{r}_2 \rvert} \omega_L(\textbf{r}_2) d\textbf{r}_1 d\textbf{r}_2
\end{equation}
where $L$ indexes the auxiliary basis function. 
The two-center two-electron repulsion integrals are also calculated as 
\begin{equation}
\label{eqn: two_center_AO_integrals}
    (\omega_{\alpha} \vert \omega_{\beta}) = \int \dfrac{ \omega_{\alpha}(\textbf{r}) \omega_{\beta}(\textbf{r}') }{ \lvert \textbf{r} - \textbf{r}' \rvert } d\textbf{r} d\textbf{r}'
\end{equation}
The three-index tensor of the three-center integrals is contracted with the density matrix by performing element-wise multiplication over corresponding indices of the atomic orbitals, followed by summation, to produce $\textbf{j} = \left\{\sum (\alpha \beta \vert l) \cdot \bm{\Gamma}_{\alpha \beta} \right\}_{l=1}^M$, the projection of the Coulomb potentials on the auxiliary basis functions. Then, $\textbf{p}$ is obtained by solving $( \omega_{\alpha} \vert \omega_{\beta} ) \cdot \textbf{j} = \textbf{p}$, a system of linear equations. Finally, the guess of $\mu_{\textnormal{opt}}$ is made from the calculation of the total kinetic energy with the summation of the one-electron kinetic energy integrals from the atomic orbitals as 
\begin{equation}
\label{eqn: kinetic_energy_integrals}
    \textbf{T}_{\alpha \beta} = -\int \dfrac{1}{2} \phi_{\alpha}(\textbf{r}) \nabla^2 \phi_{\beta}(\textbf{r}) d\textbf{r} 
\end{equation}
followed by the optimization of $\mu$ with \autoref{eqn: mu_optimization} using the initialized electron density. We note our outlined procedure can be extended to any spin configuration by considering $\textbf{p}^\alpha, \textbf{p}^\beta$, the respective vectors of density coefficients for the $\alpha, \beta$-spin densities.

With $\textbf{p}$ obtained, its elements can be iteratively optimized to optimize the electron density computed from $\bm{\Phi} \cdot \textbf{p}$, the dot product between the atomic orbital function values and the density coefficients. We perform this optimization with gradient descent, which has also been applied to KSDFT~\citep{yoshikawa2022automatic} and machine-learned OFDFT workflows~\citep{zhang2024overcoming, remme2025stable}. Our loss function includes the total energy computed at each density distribution with the PBE exchange-correlation energy functional and with the $\mu_{\textnormal{opt}}$ guess being fixed throughout each $k$ gradient descent step. 
To that total energy loss, denoted by $\mathcal{L}_{E[n]} = E[n]$, we add two additional loss functions motivated by Kato's cusp condition theorem~\citep{kato1957eigenfunctions, march1986spatially} and the virial theorem \citep{georgescu1999virial, weislinger1974classical, mclellan1974virial}. Kato's theorem states that for generalized Coulomb potentials, the electron density exhibits a cusp at a nucleus of charge $Z$ at position $\textbf{R}$ satisfying
\begin{equation}
\label{eqn: katos_cusp_condition} 
    Z = -\dfrac{1}{2 \rho(\textbf{r})} \left. \dfrac{d \rho(\textbf{r})}{dr} \right \vert_{r \rightarrow \textbf{R}}
\end{equation}
A recent work~\citep{hutcheon2024exact} has already shown that the inclusion of the cusp condition as a loss function during a SCF OFDFT calculation with the Thomas-Fermi~\citep{thomas1927calculation, enrico1927metodo} and the von Weizs\"{a}cker~\citep{weizsacker1935theorie} functionals produces qualitatively-reasonable electron densities that exhibit the cusp condition and shell-structures near nuclei. We estimate the deviation of the density from the cusp condition around a given nucleus, $Z_i$, with the loss function used in that work, 
\begin{equation}
\label{eqn: cusp_condition_loss}
    \mathcal{L}_{Z_i} = \int \left| \dfrac{\lvert \nabla \rho(\textbf{r}) \rvert^2}{8 \rho(\textbf{r})} - \dfrac{Z^2 \rho(\textbf{r})}{2} \right| w_i(\textbf{r}) d\textbf{r}
\end{equation} 
This loss function estimates the deviation from the von Weizs\"{a}cker energy that is exact for single-orbital systems and can be approximated as being true near the nucleus with the radially-decaying indicator function
\begin{equation}
\label{eqn: nucleus_indicator}
    w_i(\textbf{r}) = e^{-Z_i \textbf{r}} (\pi / Z_i^3)^{1/2}
\end{equation}
We then set our second loss as 
\begin{equation}
\label{eqn: cusp_condition_loss}
    \mathcal{L}_{Z} = \sum_{i=1}^{N_{\textnormal{atom}}} \mathcal{L}_{Z_i}
\end{equation}

We introduce a third loss term to enforce the quantum virial relation of $2 \langle \bm{\hat{\mathbf{T}}} \rangle = -\langle \bm{\hat{\mathbf{V}}} \rangle$ where $\bm{\hat{\mathbf{T}}}$ and $\bm{\hat{\mathbf{V}}}$ are respectively the kinetic energy and potential energy operators, with the latter consisting of the external, Hartree, and exchange-correlation potentials:
\begin{equation}
\label{eqn: virial_relation_loss}
    \mathcal{L}_{\textnormal{virial}} = \lvert 2\mathbf{T}_{\textnormal{total}} + \mathbf{V}_{\textnormal{total}} \rvert
\end{equation}

We optimize $\textbf{p}$ by minimizing MuAPBEK, our custom functional that we formulate as a loss function 
\begin{equation}
\label{eqn: loss_function}
    \mathcal{L}_{\text{MuAPBEK}} = \mathcal{L}_{E[n]} + \mathcal{L}_{Z} + \mathcal{L}_{\textnormal{virial}}
\end{equation}
Since $\mathcal{L}_Z$ and $\mathcal{L}_{\textnormal{virial}}$ are already in terms of energies expressed in atomic units, we reason that they do not require scalar or dimensional multipliers.
The partial derivatives in the loss gradient $\nabla_{\textbf{p}} \mathcal{L}$ are computed via automatic differentiation using \texttt{PyTorch}~\citep{paszke2019pytorch}. 
To conserve the number of electrons, $N_e$, divergence from the hyperplane of normalized densities is prevented as in \cite{zhang2024overcoming} by projecting $\nabla_{\textbf{p}} \mathcal{L}$ onto the tangent space of normalized densities, 
\begin{equation}
\label{eqn: tangent_space_projection} 
    \nabla_{\textbf{p}} \mathcal{L}^{\textnormal{proj.}} = \left( \textbf{I} - \frac{\textbf{w} \textbf{w}^\top}{\textbf{w}^\top \textbf{w}} \right) \nabla_{\textbf{p}} \mathcal{L}
\end{equation}
where $\textbf{w}$ is a vector of density basis function integrals with $\textbf{w}_{\alpha} = \int \omega_{\alpha}(\textbf{r}) d\textbf{r}$ as its elements. 
For stability during gradient descent, we further normalize $\nabla_{\textbf{p}} \mathcal{L}^{\textnormal{proj.}}$ as $\nabla_{\textbf{p}} \mathcal{L}^{\textnormal{proj.}}_{\textnormal{norm.}}$ and update the density coefficients as
\begin{equation}
\label{eqn: density_coefficients_update} 
    \textbf{p}^{(k + 1)} = \textbf{p}^{(k)} - \epsilon \cdot \nabla_{\textbf{p}} \mathcal{L}^{\textnormal{proj.}}_{\textnormal{norm.}}
\end{equation}
where $\epsilon$ is the learning rate.

\begin{table}[!t]
\begin{center}
\caption{\label{table: single_atom_performances}
OFDFT signed errors ($E_{\rm h}$) to PBE/def2-tzvp KSDFT results on single atoms of charge $3\leq Z \leq 36$.
``APBEK\texttt{+}" indicates density optimization via minimization of only $\mathcal{L}_{E[n]}$ but with the $\mu_{\textnormal{opt}}$ guess. 
MuAPBEK achieves much lower errors than APBEK and APBEK\texttt{+} due to its constraints of the cusp condition and virial theorem. 
}
\begin{tabular}{lccc}
\textbf{$Z$} & \textbf{APBEK} & \textbf{APBEK\texttt{+}} & \textbf{MuAPBEK}\\ 
\hline 
Li & -0.129 & -0.127 & 0.001\\
Be & -0.056 & -0.056 & -0.047\\
B & -0.587 & -0.581 & 0.003\\
C & -0.766 & -0.748 & 0.010\\
N & -0.831 & -0.804 & 0.023\\
O & -0.948 & -0.923 & -0.012\\
F & -1.065 & -1.048 & -0.012\\
Ne & -0.208 & -0.207 & -0.114\\
Na & -1.128 & -1.133 & 0.012\\
Mg & -0.153 & -0.154 & -0.065\\ 
Al & -1.140 & -1.156 & 0.011\\
Si & -1.094 & -1.112 & -0.010\\
P & -1.362 & -1.387 & 0.014\\
S & -1.360 & -1.385 & -0.025\\
Cl & -1.345 & -1.372 & -0.098\\
Ar & -0.162 & -0.162 & -0.060\\
K & -1.150 & -1.173 & -0.266\\
Ca & -0.152 & -0.154 & -0.172\\
Sc & -0.788 & -0.791 & 0.538\\
Ti & -1.293 & -1.293 & 0.175\\
V & -1.140 & -1.135 & 0.313\\
Cr & -1.262 & -1.250 & 0.153\\
Mn & -3.737 & -3.743 & -1.507\\
Fe & -4.095 & -4.098 & -1.702\\
Co & -3.957 & -3.959 & -1.274\\
Ni & -3.498 & -3.501 & -0.220\\
Cu & -1.330 & -1.318 & 0.936\\
Zn & -0.268 & -0.269 & -0.095\\
Ga & -1.840 & -1.840 & 0.106\\
Ge & -1.717 & -1.720 & 0.093\\
As & -1.958 & -1.968 & 0.066\\
Se & -1.824 & -1.837 & 0.067 \\
Br & -1.677 & -1.693 & 0.069\\
Kr & -0.513 & -0.511 & 0.061\\
\hline
\textbf{MAE} (kcal/mol) & \textbf{821} & \textbf{823} & \textbf{161}\\
\hline
\end{tabular}
\end{center}
\end{table}

\begin{figure}[t]
\resizebox{\columnwidth}{!}{\includegraphics{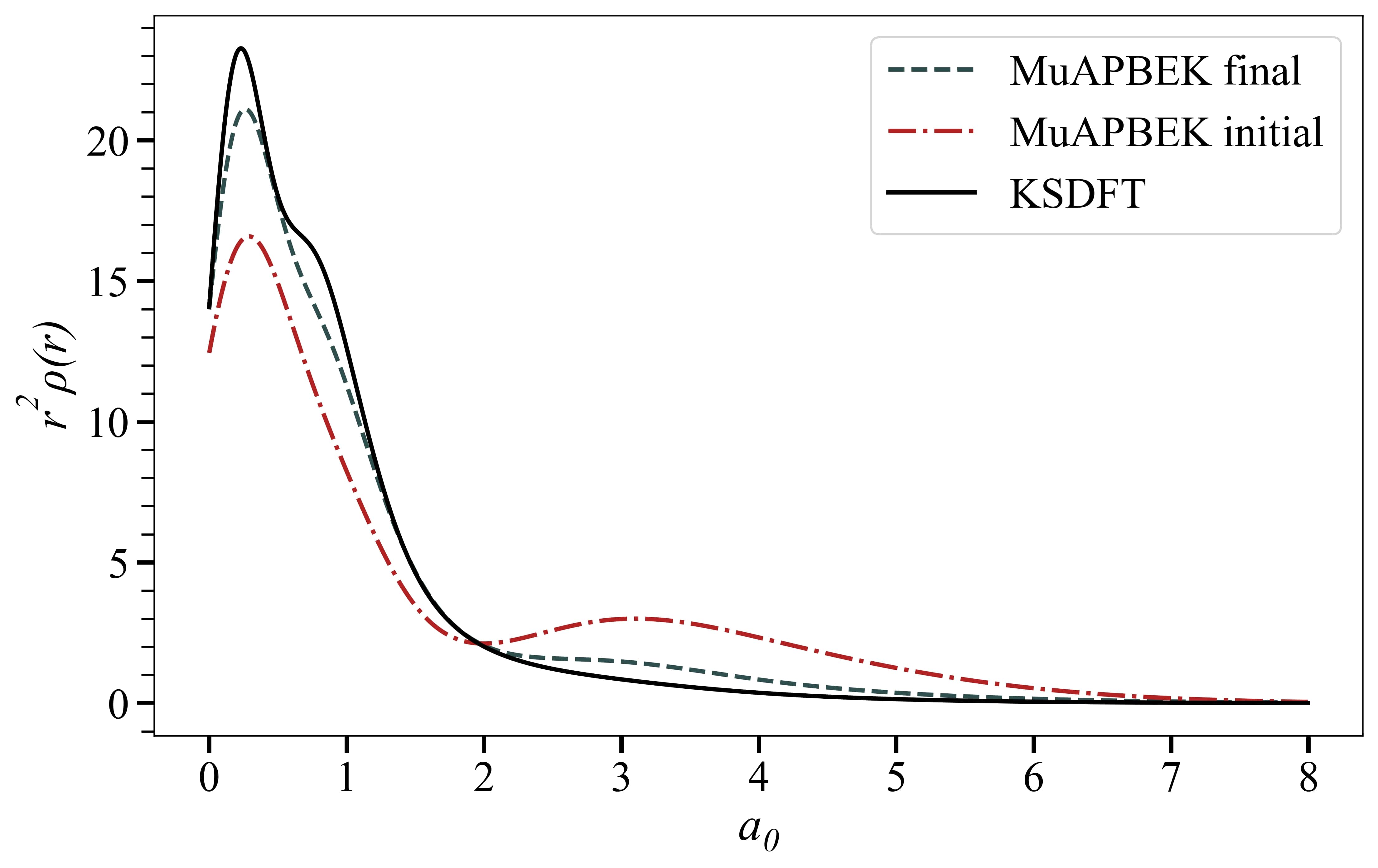}} 
\caption{\label{fig: Fe_atom_electron_density_distribution}
Radial electron density distributions of iron atom centered about its nucleus. Despite the poor initial density, MuAPBEK produces a final density that is in closer agreement with the KSDFT density by satisfying the cusp condition, having a more pronounced peak near the core, and decaying more smoothly in the tail.
}
\end{figure} 

We apply our workflow to single atoms with $3\leq Z \leq 36$ in their ground-state spin configurations using a stochastic gradient descent (SGD) optimizer with $\epsilon = 0.001$ and momentum of 0.9 for $k=100$. We have not analyzed hydrogen and helium since the von Weizs\"{a}cker functional is already exact for these two atoms. We also leave the analysis of heavier atoms, for which considerations such as relativistic effects become important \citep{pyykko2012relativistic}, to future works. The def2-universal-JK-fit auxiliary basis set and PBE exchange-correlation energy functional are employed. These hyperparameters enable rapid convergence of $\mathcal{L}$, suggesting the effective identification of the Pareto front among the different loss functions. From this optimization, we select \textbf{p} that minimizes $E$, which gives the results benchmarked against PBE/def2-tzvp KSDFT in \autoref{table: single_atom_performances}. MuAPBEK achieves a mean absolute error, MAE, of 161 kcal/mol in total energies. Notably, this is a five-fold improvement over APBEK when for sake of comparison, it was also optimized for $k=100$, although its error would likely be much worse at full convergence. We find that tuning $\mu$ alone is insufficient in obtaining accurate energies and densities since ``APBEK\texttt{+}", containing the $\mu_{\textnormal{opt}}$ guess but lacking $\mathcal{L}_{Z}$ and $\mathcal{L}_{\textnormal{virial}}$, performs comparably to APBEK. We also observe that the largest errors occur for the transition metals with poor initial density guesses. For example, MuAPBEK gives its worst result for iron, with an error of -1.702 $E_{\textnormal{h}}$, which has a significantly inaccurate starting density (\autoref{fig: Fe_atom_electron_density_distribution}). Encouragingly, MuAPBEK still manages to improve upon this initial guess by capturing the correct nuclear cusp and a smoother decay in the tail-region. These findings underscore the importance of having a high-quality initial density in our OFDFT calculations to ensure the density converges to the correct distribution with an accurate energy prediction.

\begin{figure}[!t]
\begin{center}
\resizebox{\columnwidth}{!}{\includegraphics{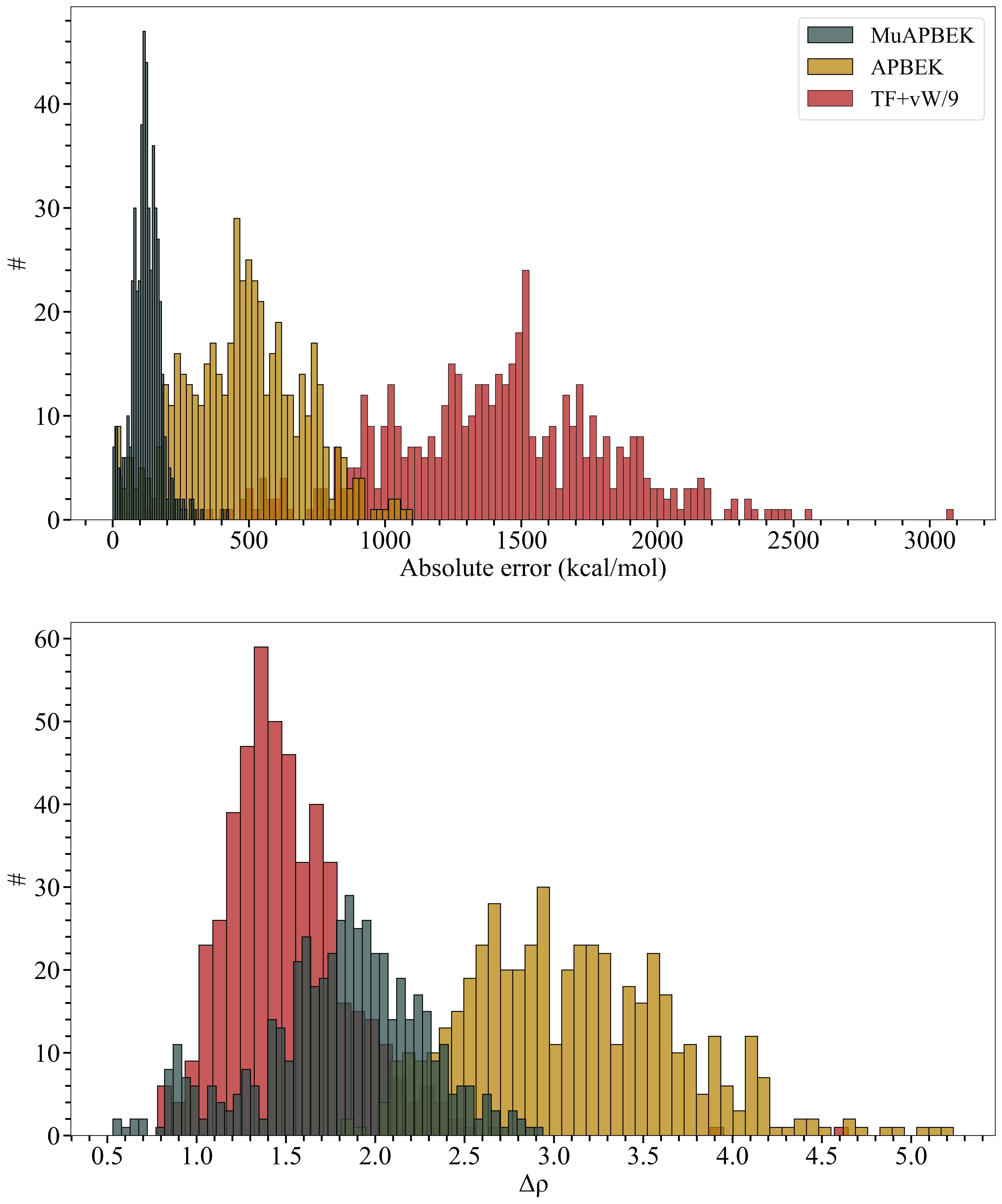}} 
\caption{\label{fig: QM9_results_optimized}
\textbf{Top:} Distributions of total energy absolute errors (kcal/mol) of MuAPBEK, APBEK, and TF+vW/9 relative to PBE/def2-qzvp KSDFT values for 500 random QM9 molecules. 
MuAPBEK's MAE is 122 kcal/mol, which is much lower than APBEK's and TF+vW/9's MAEs of 482 and 1449 kcal/mol, respectively.
\textbf{Bottom:} Distribution of integrated absolute differences (\autoref{eqn: integrated_density_difference}) between the optimized OFDFT and KSDFT densities for the same 500 molecules. 
MuAPBEK's density error is 1.8 electrons, and APBEK's and TF+vW/9's are 3.1 and 1.6 electrons, respectively. 
}
\end{center}
\end{figure} 

We also compute the total energies of 500 randomly-selected molecules in QM9 (\autoref{fig: QM9_results_optimized}, top). 
Relative to PBE/def2-qzvp KSDFT, MuAPBEK achieves a MAE of 122 kcal/mol from its optimized densities. 
This error is of similar magnitude to what was obtained for the single atoms, which suggests that the non-empirical formulation of MuAPBEK grants it strong generalization capabilities. 
In contrast, APBEK and the classic kinetic energy functional TF+vW/9 (the Thomas-Fermi functional with the von Weizs\"{a}cker gradient correction~\citep{benguria1981thomas}) produce significantly higher MAE values of 482 and 1449 kcal/mol, respectively, when they were optimized for $k = 100$ steps. 
Even when they are evaluated on converged KSDFT densities, their MAE values are 332 and 1632 kcal/mol, respectively. 
MuAPBEK's densities also align well with KSDFT's, with a mean absolute difference in the integrated densities for the molecules, with an average of 66 electrons, of 1.8 electrons (\autoref{fig: QM9_results_optimized}, bottom), calculated as 
\begin{equation}
\label{eqn: integrated_density_difference}
    \Delta \rho = \int \left\lvert \rho(\textbf{r})^{\textnormal{OFDFT}} - \rho(\textbf{r})^{\textnormal{KSDFT}} \right\rvert d\textbf{r}
\end{equation} 
APBEK sees its densities diverge as its mean density error is 3.1 electrons. 
Interestingly, TF+vW/9 appears to benefit from the density optimization as its mean density error of 1.6 electrons is comparable to MuAPBEK's. 
However, its energy errors are clearly much too large and suggest that the family of Thomas-Fermi-von Weizs\"{a}cker functionals are a poor baseline to which corrections can be made, as we have done in this work for APBEK.

\begin{figure}[!t]
\resizebox{\columnwidth}{!} 
{\includegraphics{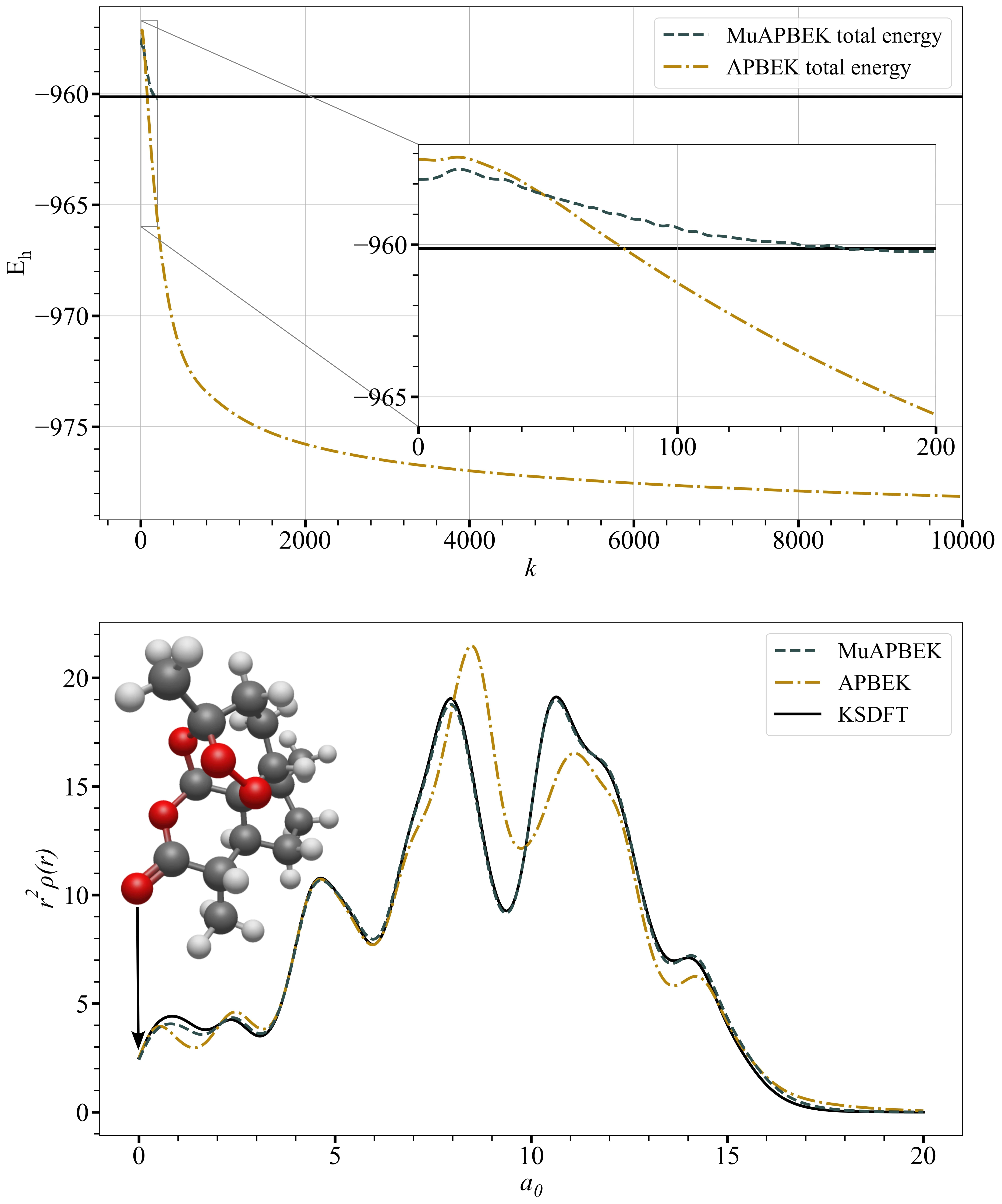}} 
\caption{\label{fig: Artemisinin}
\textbf{Top}: Evolutions of total energies of MuAPBEK and APBEK with $k$ against PBE/def2-qzvp KSDFT result (wall time of 71 minutes) on the anti-malarial drug artemisinin, shown with the horizontal line. 
MuAPBEK converges within 200 steps (6 minutes) with an absolute total energy error of 62 kcal/mol, whereas APBEK has not converged even after 10000 steps (325 minutes) to an error of 11,291 kcal/mol. 
\textbf{Bottom:} Radial electron density distributions centered about the double-bonded oxygen in the ester group. 
The absolute integrated density errors for MuAPBEK and APBEK are 2.9 and 19.3 electrons, respectively.
}
\end{figure} 

We further assess MuAPBEK's generalization capabilities to a molecule larger than those found in QM9. 
Indeed, we have analyzed artemisinin \citep{klayman1985qinghaosu, white2008qinghaosu, tu2011discovery}, a plant-derived organic drug with seventeen heavy atoms (thirty-nine atoms, total) used worldwide to treat malaria. Relative to PBE/def2-qzvp KSDFT, MuAPBEK produces a total energy prediction that is accurate to within 62 kcal/mol after running it to convergence for $k=200$ steps (\autoref{fig: Artemisinin}, top). Its resulting density distribution is also in good agreement with that of KSDFT (\autoref{fig: Artemisinin}, bottom), as the absolute integrated density error is 2.9 electrons. 
The speed advantage of MuAPBEK is also evident since on identical hardware (16-core, 4.5 GHz AMD Ryzen 9 7950X CPU and 128 GB of DDR5 RAM), MuAPBEK finished in 6 minutes whereas KSDFT took 71 minutes. 
For comparison, we also attempted to converge APBEK by running it for $k=10,000$ steps (\autoref{fig: Artemisinin}, top). However, we found that APBEK's total energy was still decreasing at $k=10,000$, by which point it yielded an absolute total energy error of 11,291 kcal/mol and a density error of 19.3 electrons (\autoref{fig: Artemisinin}, bottom). Although our analysis was limited to a single molecule, these results on artemisinin along with those on QM9 demonstrate that MuAPBEK may enable a critical advance toward fast and reliable OFDFT simulations of realistic chemical and material systems at sizes larger than those that are feasible with routine KSDFT methods. If this is realized, the search for \textit{e.g.} novel and potent medical drugs could become much more efficient.

\begin{figure}[!t]
\resizebox{\columnwidth}{!}{\includegraphics{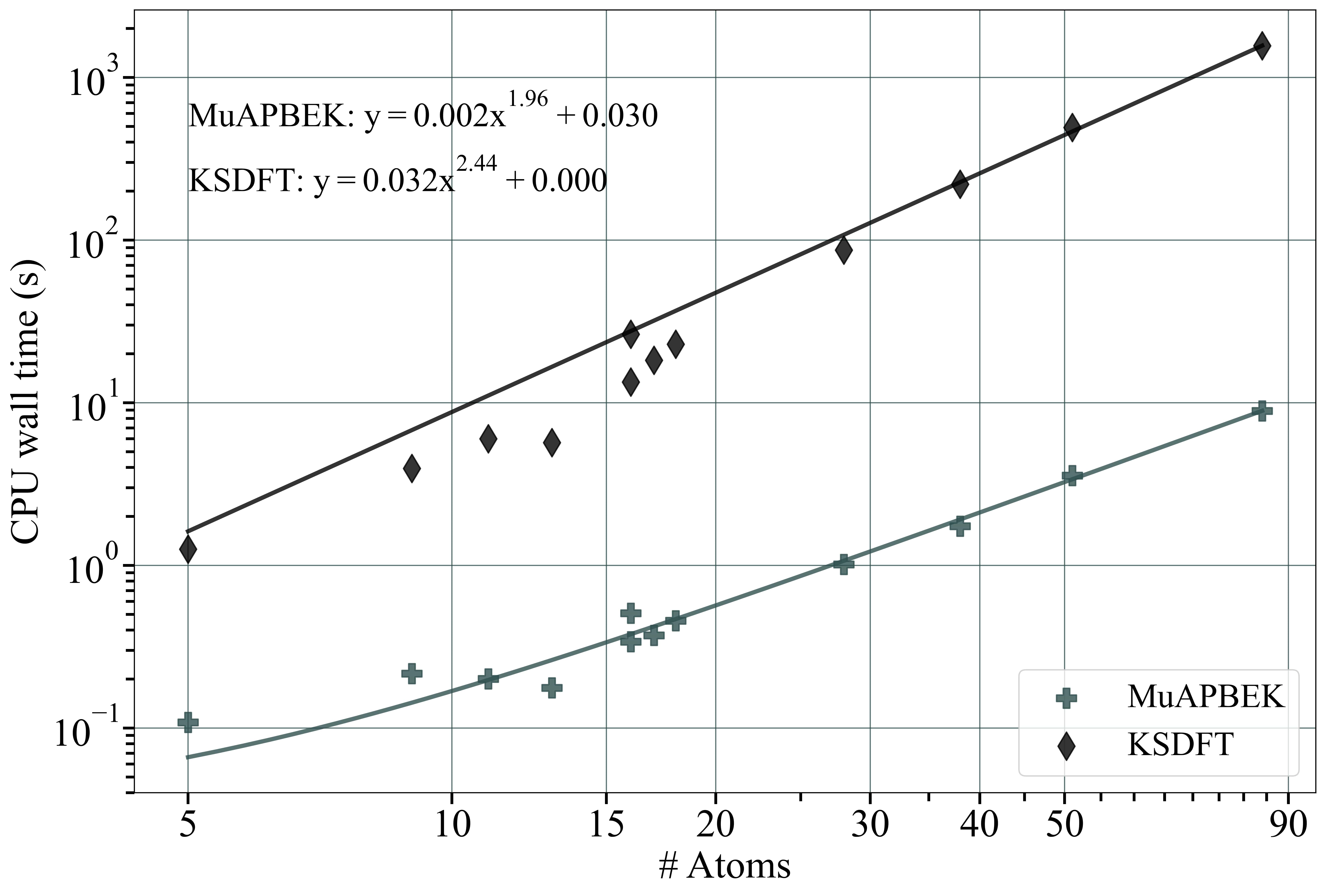}} 
\caption{\label{fig: Computational_complexity}
CPU wall times of MuAPBEK and KSDFT for a single density optimization step as a function of total number of atoms timed on equal computer hardware. 
MuAPBEK scales as $\mathcal{O}(N^{1.96})$, which is of lower order than the observed $\mathcal{O}(N^{2.44})$ KSDFT scaling. 
}
\end{figure} 

Finally, we empirically demonstrate the scaling advantage of MuAPBEK over KSDFT. From the QMugs dataset~\citep{isert2022qmugs}, which includes pharmacologically-relevant molecules with up to 100 heavy atoms, we randomly selected twelve molecules spanning a range of sizes and measured the average time per density optimization step for both KSDFT and MuAPBEK. For this analysis, all computations were performed on the aforementioned hardware. As shown in \autoref{fig: Computational_complexity}, MuAPBEK always achieves a lower wall time. For the smallest molecule consisting of three heavy atoms (five atoms, total), MuAPBEK and KSDFT were timed at 0.11 seconds and 1.26 seconds, respectively. For the largest molecule consisting of fifty heavy atoms (eighty-four atoms, total), the wall times were 9 seconds and 1565 seconds, respectively. MuAPBEK exhibits $\mathcal{O}(N^{1.96})$ scaling that is of lower order than the observed $\mathcal{O}(N^{2.44})$ scaling for KSDFT. This promises orders-of-magnitude faster calculations for systems much larger than those that routine KSDFT calculations can afford. This favorable quadratic scaling arises due to the $\mathcal{O}(M^2)$ time complexity of the matrix-vector multiplication involving the two-center two-electron integral matrix and density coefficients in the Hartree energy evaluation. However, our implementation requires the evaluations of the kinetic and exchange-correlation energies on real-space grid points, which number on the order of $\sim$$10^4 N_{\textnormal{atom}}$. This would pose a serious challenge of loading these grid points onto memory for large systems, especially at the high floating-point precision standards common to quantum chemistry \citep{von2021arbitrarily, laqua2021accelerating, dawson2024reducing}.

In conclusion, we demonstrate that the APBEK kinetic energy functional can be efficiently parameterized by fitting its $\mu$ parameter to the total kinetic energy obtained from the one-electron kinetic energy integrals of the initialized electron density. By augmenting this parameterized functional with two energy functionals derived from Kato's cusp condition and the virial theorem, and minimizing this resulting functional, which we call MuAPBEK, we achieve significantly improved energies and densities compared to standard analytical kinetic energy density functionals like APBEK and TF+vW/9. MuAPBEK's MAE is consistent across both atoms and molecules, indicating strong size-independence and generalization. 
It also delivers at least ten-time speedups over a KSDFT SCF cycle and scales favorably with a time complexity of $\sim$$\mathcal{O}(N^2)$. These results point to a clear, promising route toward reliable, routine OFDFT computations in \textit{e.g.} the exploration of chemical and material spaces, and large-scale simulations of biological systems. Future work may continue non-empirical kinetic energy functional development to reduce the energy and density errors we have achieved to near-chemical accuracy. 
One avenue may be to consider other enhancement factors while further assessing the effectiveness of the conjointness conjecture. In this spirit, our proposed method of tuning $\mu$ in our work using the kinetic energy integrals from the initialized density may be generalized to tune a given parameter in any kinetic energy functional. For example, the $\kappa$ parameter in \autoref{eqn: PBE_exchange_enhancement_factor} may be tuned independently from or simultaneously with $\mu$ to perhaps yield improved results. Additionally, due to our current requirement of a fine real-space grid, it would be worthwhile to devise a more memory-efficient method, such as through efforts to develop more compact meshes and to evaluate energies directly from density coefficients.

\subsection*{Acknowledgements}
Siwoo Lee appreciates helpful discussions with O. Anatole von Lilienfeld, Danish Khan, Kevin Kurian Thomas Vaidyan, and Tiara Safaei. Adji Bousso Dieng acknowledges support from the National Science Foundation, Office of Advanced Cyberinfrastructure (OAC): \#2118201. She also acknowledges Schmidt Sciences for the AI2050 Early Career Fellowship and DataX.

\bibliographystyle{apa}
\bibliography{References}

\end{document}